%% file: radion_v4.tex
\documentclass[aps,prl,reprint,groupedaddress,nofootinbib]{revtex4-1}
\usepackage{hyperref}
\usepackage{amsmath}
 \usepackage{multirow}
\usepackage{array}
\newcolumntype{L}[1]{>{\raggedright\let\newline\\\arraybacksslash\hspace{0pt}}m{#1}}
\newcolumntype{C}[1]{>{\centering\let\newline\\\arraybackslash\hspace{0pt}}m{#1}}
\newcolumntype{R}[1]{>{\raggedleft\let\newline\\\arraybackslash\hspace{0pt}}m{#1}}
\usepackage{float}
\usepackage{graphicx,epstopdf}
\usepackage{epsfig}
\usepackage{subfigure} 
\usepackage{psfrag}
\usepackage{color}  
\usepackage{slashed}

\usepackage{amsfonts}  
\usepackage{bbm}
\usepackage{bm}
\usepackage{amssymb}
\usepackage{tikz}
\usepackage{tikz}
\usetikzlibrary{positioning,arrows}
\usetikzlibrary{decorations.pathmorphing} 
\usetikzlibrary{decorations.markings}
\usepackage{flushend}

\newcommand*{\be}{\begin{equation}}
\newcommand*{\ee}{\end{equation}}

\newcommand{\comment}[1]{}

\newcommand{\cref}[1]{Chapter~\ref{c.#1}}

\newcommand{\drawsquare}[2]{\hbox{%
\rule{#2pt}{#1pt}\hskip-#2pt
\rule{#1pt}{#2pt}\hskip-#1pt
\rule[#1pt]{#1pt}{#2pt}}\rule[#1pt]{#2pt}{#2pt}\hskip-#2pt
\rule{#2pt}{#1pt}}

\newcommand{\Yfund}{\raisebox{-.5pt}{\drawsquare{6.5}{0.4}}}

\def\beq{\begin{eqnarray}}
\def\eeq{\end{eqnarray}}
\def\ba{\begin{array}}
\def\ea{\end{array}}
\def\bi{\begin{itemize}}
\def\ei{\end{itemize}}
\def\be{\begin{enumerate}}
\def\ee{\end{enumerate}}
\def\bc{\begin{center}}
\def\ec{\end{center}}
\def\bt{\begin{table}}
\def\et{\end{table}}
\def\btb{\begin{tabular}}
\def\etb{\end{tabular}}

\def\lsim{\raise0.3ex\hbox{$\;<$\kern-0.75em\raise-1.1ex\hbox{$\sim\;$}}}
\def\gsim{\raise0.3ex\hbox{$\;>$\kern-0.75em\raise-1.1ex\hbox{$\sim\;$}}}

\usepackage[normalem]{ulem}

\begin{document}

\title{Radion dynamics in  the Multibrane Randall-Sundrum Model}

\author{Haiying Cai$^{1}$}
\email{hcai@korea.ac.kr}
\affiliation{
$^1$Department of Physics, Korea University, Seoul 136-713, Korea
}

\begin{abstract} 
The radion equilibrium in the Randall-Sundrum  (RS) model is guaranteed by  the  backreaction of a bulk scalar field. In this paper we study an extended scenario,  where an intermediate  brane exists in addition to the two branes  at the  fixed points,  due to the discontinuity of bulk cosmology constants in  two spatial regions. We conducted a complete  analysis of the  linearized Einstein's equations after applying the Goldberger-Wise mechanism.  Our result elucidates that  in the presence of nonfixed point branes under the rigid assumption, a unique radion field is conjectured as legitimate in the RS metric perturbation.  The cosmological expansion in this setup is briefly discussed.
\end{abstract}

\maketitle

Extra dimension theories have been constructed to resolve the gauge hierarchy problem without invoking the supersymmetry~\cite{Horava:1995qa, Arkani-Hamed:1998jmv, Antoniadis:1998ig}.  One theoretical appealing proposal is  a slice of  anti-de Sitter (AdS$_5$) space as a solution to Einstein's equation with  a negative bulk cosmology constant plus  opposite brane tensions~\cite{Randall:1999vf, Randall:1999ee}. This  RS model can  naturally explain the weakness of  graviton coupling to SM matters and  the TeV  scale  emerges  from a warped geometry factor.  To achieve the expected  scale at the infrared brane,  a dimensionless parameter should be $k L \sim 35$, with the curvature $k = \sqrt{-\frac{\Lambda}{12 M^3}}$, in terms of  the 5D Planck scale and bulk cosmology constant.   Thus as an attempt to address the hierarchy problem, the brane separation needs to be stabilized at $L \sim 35/k $.  An observation in light of  the late-time cosmology in brane models also calls for  stabilizing the flat direction. For the RS$1$,   the Einstein equation $G_{55} = \kappa^2 T_{55}$  evaluated  at the brane predicts $H^2 \sim  \rho\left(2 \Lambda_w + \rho \right)$, with the quadratic term $\Lambda_w^2$ cancelled by the negative $\Lambda$ in the bulk~\cite{Binetruy:1999ut}.  Without stabilizing  the radius,  in the IR brane,   it seems that the matter and radiation density is forced to be negative ($\rho < 0$) for $\Lambda_w <0$~\cite{Cline:1999ts}. While this constraint can be eliminated after including the effect of radion stabilization~\cite{Csaki:1999mp, Kanti:1999sz}.  As a consequence, the radion acquires a mass and the conventional Friedmann-Robertson-Walker (FRW) equations are  recovered.  

One elegant way to stablise the radius was the Goldberger-Wise (GW) mechanism~\cite{Goldberger:1999uk, Goldberger:1999un}, by introducing a massive bulk scalar minimally coupled to the graviton. With appropriate  brane terms,  the bulk scalar can develop a $y$-dependent vacuum expectation value (VEV), so that the effective potential of radion  after integration  over the fifth dimension will gain an extrema at a fixed brane separation.  In this letter, we  are going to  investigate  the radion dynamics after imposing GW mechanism in  a  RS model   with multibranes~\cite{Kogan:2001qx}. The authors in that paper worked out  $2$ spin-$0$ radion  fields  in the metric perturbation,  with one of them representing the relative moving of  the intermediate brane. However, that claimant contradicts the degrees of freedom counting in a $5 D$ extra-dimension theory~\cite{Giudice:1998ck, Han:1998sg}.  We will illustrate  below  that assuming the intermediate brane is {\it{rigid}} for a static configuration, only one radion field is  the legitimate solution to  the  linearized Einstein's equations with the {\it{jump }}conditions matched at  all  branes. Following that proof, the radion mass will be derived by including the backreaction of the GW bulk scalar.

We  start with the metric ansatz $ g_{MN} dx^M dx^N$ perturbed by the transverse graviton $h_{\mu \nu}(x,y)$ and  one radion field $f(x)$ on an $S^1/Z_2$ orbifold:
\beq
d s^2  &=&   e^{-2 A (y) - 2 F(y) f (x) }\left[ \eta_{\mu \nu} + 2 \epsilon(y) \partial_{\mu} \partial_{\nu} f (x) \right. \nonumber \\ 
&+& \left. h_{\mu \nu}(x,y)\right] d x^\mu d x^\nu - \left[1+ G(y) f(x) \right]^2 dy^2 \,, \label{metric} 
\eeq
where the subscripts are $\mu,\nu =0,1,2,3$ and $M,N \in (\mu, 5)$. Compared with the metric in~\cite{Csaki:2000zn},  the fluctuation of $ 2 \epsilon(y) \partial_{\mu} \partial_{\nu} f (x)$  is assumed to be permitted.  Now we consider the five dimensional action of graviton coupling to a single bulk  scalar field:
\beq
& - & \frac{1}{2 \kappa^2} \int d^5 x \sqrt{g} \, {\cal R} + 
\int d^5 x \sqrt{g}  \Big( \frac{1}{2} g^{M N}  \partial_M \phi \partial_N \phi
-  V(\phi) \Big)  \nonumber \\ &-& \int d^5x\frac{\sqrt{g}}{\sqrt{-g_{55}}}\sum_{i} \lambda_{i}(\phi) \delta(y - y_i)
\,,\label{Act}
\eeq
with $\kappa^2 = 1/(2 M^3)$ and $y_i = \{0, \pm r , L \} $  designating the location of  branes in the $y$-coordinate. Note that the action integration spans over the entire $S^1$ circle  with $ L>r >0$. The bulk scalar can be expanded around a $y$-dependent vacuum expectation value (VEV): $\phi(x,y) = \phi_0(y) + \varphi(x,y)$. Note that the brane terms are crucial to compensate the discontinuity caused by the orbifold compactification. By varying the action Eq.(\ref{Act}) with respect to the 5d metric tensor $g_{MN}$, one can derive the Einstein's equations in terms of Ricci tensor: $R_{MN} = \kappa^2 \tilde{T}_{MN} \equiv \kappa^2 \left(T_{MN} -\frac{1}{3} g_{MN} T^a_a\right)$, with  the energy-momentum tensor  given by $T_{M N} = 2 \delta \left(  \sqrt{g}\, \mathcal{L}_\phi \right) / \left( \sqrt{g} \, \delta g^{MN} \right)$.  It is advantage to work in this approach because  at most the linear order perturbation is involved. We  have calculated the Ricci tensor  $R_{MN}$ from the metric,  and the source term  $\tilde{T}_{MN}$  can be written in a compact form:
 \beq
  \tilde{T}_{\mu \nu} &=&  -\frac{2}{3} g_{\mu \nu} V(\phi) -\frac{1}{3} \sqrt{-g^{55}} g_{\mu \nu} \sum_i  \lambda_i (\phi)\delta(y - y_i)
 \nonumber \\ 
 \tilde{T}_{\mu 5} &=&   \partial _{\mu} \phi \partial_5 \phi  \\
  \tilde{T}_{55} &=&  (\partial_5 \phi)^2  - \frac{2 g_{55}}{3}  V(\phi) + \frac{4  }{3}  \sqrt{-g_{55}} \sum_i  \lambda_i (\phi) \delta(y - y_i) \nonumber 
 \eeq
 By decomposing $R_{MN}$ and $\tilde{T}_{M N}$  till  the linear order of metric perturbations, we listed their exact expressions (Eq.(\ref{Ruv}-\ref{T55})) in  Appendix~\ref{Appendix}.

To the zeroth order,  one can obtain the background  (BG) equations  for the VEV $\phi_0$ and metric $A$:
\beq
&&\phi_0 '' = 4 A' \phi_0 ' + \frac{\partial V(\phi_0)}{\partial \phi}+ 
\sum_i \frac{\partial \lambda_i(\phi_0)}{\partial \phi} \delta(y-y_i)\,, \\
&& 4 A'^2-A'' = -\frac{2 \kappa^2 }{3}  V(\phi_0) - \frac{\kappa^2 }{3}
\sum_i \lambda_i(\phi_0) \delta(y-y_i) \,, \label{BG1}
\\
&& A'^2 = \frac{\kappa^2 {\phi_0}'^2}{12} 
- \frac{\kappa^2}{6} V(\phi_0) \,.
\label{BG2}
\eeq
where the prime denotes the partial derivative with respect to $y$ and the last equation originates from $G_{55}= \left(R_{55} - \frac{1}{2}g_{55} \cal{R} \right) = \kappa^2 T_{55}$. The analytic solutions for these nonlinear equations can be found  using the superpotential method~\cite{DeWolfe:1999cp, Behrndt:1999kz}, with the  backreaction effect automatically accounted. Provided the bulk potential $V(\phi)$ can be written in the form of:
\begin{equation} 
V(\phi) = \frac{1}{8}
\left[\frac{\partial W(\phi)}{\partial \phi}\right]^2
-\frac{\kappa^2}{6} W(\phi)^2  \,,
\end{equation}
then a solution  to the BG equations is given by:
\begin{equation} 
\phi_{0} ' = \frac{1}{2} \frac{\partial W}{\partial \phi} \hbox{ , } 
A'=\frac{\kappa^2}{6} W(\phi_0)  \,. \label{solution}
\end{equation} 
To reproduce the usual exponential metric in multibrane RS model~\cite{Kogan:2001qx},  the superpotential is  derived as:
\beq
W(\phi) = \begin{cases} \frac{6 k_1}{\kappa^2} - u \phi^2 \,, &  0 < y < r  \\[0.3cm]  
\frac{6 k_2 }{\kappa^2} - u \phi^2  \,,  & r < y < L  
\end{cases}   
\eeq
with the  brane potentials:
\beq
&&\lambda_{\pm} = \pm W(\phi_\pm) \pm W'(\phi_\pm) \left(\phi - \phi_\pm \right) + \gamma_\pm \left(\phi -\phi_\pm \right)^2 \, \label{brane1}    \\
&& \lambda_{\pm r} = \frac{1}{2} \big[W(\phi(y)) \big] \big{|}_{y=r}  = \frac{3 (k_2- k_1)}{ \kappa^2} \,. \label{brane2}   
\eeq
where the subscript $\pm$ denotes the $y=(0, L)$ branes and  the {\it{jump}} for a general quantity $W$ is defined as $[W(y)]|_{y=y_i} \equiv  W(y_i+\varepsilon) - W(y_i-\varepsilon)|_{\varepsilon \to 0}$. Note that $\lambda_{\pm r}$ does not depend on the $\phi$ field since there is no {\it jump} for $\phi'(y)$ at $y=\pm r$.  We also remark that this method can be generalized to the scenario with several bulk scalars given that the superpotential is of the special class $W = \sum_{i=1}^n W_i(\phi_i)$, where each $W_i(\phi_i)$  only depends on a single scalar field. 

Now we are ready to investigate the coupled equations for the excitations using the linearized Einstein Equation $\delta R_{MN} = \kappa^2 \delta \tilde{T}_{MN}$. First of all, we need to figure out the conditions for decoupling the transverse graviton from the scalar excitation.  At the linear order, the $(\mu 5)$-component gives the first orthogonal condition:
\beq
3 \left(  F'   - A'   G \right)  \partial_\mu f (x) =   \kappa^2 \phi_0'  \partial_\mu \varphi \label{orth1} \,.
\eeq
While the $(\mu \nu)$-component is more complicated,  we can  extract out the $\partial_\mu \partial_\nu f(x) $ term from $R_{\mu \nu}$ and $\tilde{T}_{\mu \nu}$ and match them  (ref Eq.(\ref{Ruv1}-\ref{Ruv2}), Eq.(\ref{Tuv}) in Appendix{\ref{Appendix}}):
\beq
&&  e^{-2 A} \left[ 2 \left[  4 A'^2 - A'' \right]   \epsilon (y) +  \epsilon''(y) - 4 A' \epsilon'(y) \right]   +  (2 F -G)  \nonumber  \\
&& =  - \frac{2 \kappa^2 e^{-2A}   }{3} \Big(2 \epsilon(y)  V(\phi_0)  +   \sum_i \lambda_i (\phi_0)  \epsilon( y) \delta (y -y_i)  \Big) 
\eeq
The rationale is  that except for gravitons and the above ones, the other terms are all proportional to $\eta_{\mu \nu}$.  Then applying the background equation~(\ref{BG1}), we derived the second orthogonal condition:
\beq
e^{-2A} \left[ \epsilon''(y) - 4 A' \epsilon' (y) \right] + (2F -G) =0 \label{orth2}
\eeq
We would like to mention  that  Eq.(\ref{orth1}) and Eq.(\ref{orth2}) are equivalent to  the transverse and traceless gauging fixing for the graviton. In particular, Eq.(\ref{orth2}) indicates that,
 \beq
&& [\epsilon'(y)]\Large{|}_{y = \{0,  \pm r , L \}} =0 \,, \label{epc}  
 \eeq 
since there is no singular term to match here.  In analogy to the bulk graviton in the RS model that is illustrated in Appendix\ref{Appendix}~\cite{Davoudiasl:1999jd},  under the orbifold  symmetry $\epsilon'(y) = - \epsilon'(-y)$, this translates into the continuous boundary condition  $\epsilon'(0) = \epsilon'(L)= 0$ at the fixed points.  While the remaining junction condition $ [\epsilon'(y)]\Large{|}_{y =\pm r}=0$  will  constrain the integration constants of the 5d profile $F(y)$  in the two spatial regions.

To obtain the equation of motion (EOM) for the radion field,  one can construct the quantity $e^{2A} \frac{R_{\mu \nu}}{\eta_{\mu \nu}} +R_{55}$ to remove the term  of $V'(\phi_0) \varphi$ in  the  Einstein's equations. Then substituting into  that  with Eq.(\ref{BG1}) and  Eq.(\ref{orth2}) for a further simplification,  one will arrive at the following ansatz:
\beq
&& 3\left( F'' - A' G' \right) f(x) + 3 \left[ F e^{2 A} - A' \epsilon' (y) \right] \Yfund f (x)   \label {eom}\\
&=& 2 \kappa^2 \phi_0' \varphi' + \frac{\kappa^2}{3} \sum_i \left[ 3 \lambda_i (\phi_0)  G f (x)  +  3 \frac{\partial \lambda_i}{\partial \phi} \varphi \right] \delta(y - y_i) \nonumber 
 \eeq
The  discontinuity conditions of $F'$ at the branes can be obtained by  matching the singular term in the above equation:
 \beq
\left[ F'  f(x) \right] |_i =\frac{\kappa^2}{3} \left( \lambda_i G(y) f(x) +  \frac{\partial \lambda_i}{\partial \phi} \varphi (x, y)  \right) \,. \label{FBC}
\eeq
After identifying  the  ones  of  $A'$ and $\phi_0'$ at the junctions:
\beq
\left[ A'   \right] |_i = \frac{\kappa^2}{3} \lambda_i \left(\phi_0 \right) \,, ~~~ \left[ \phi_0'  \right] |_i =  \frac{\partial \lambda_i}{\partial \phi} (\phi_0) \label{ABC}
\eeq
we can see that the jump equation (\ref{FBC}) is consistent with the first orthogonal condition~(\ref{orth1}). 

Equipped with  the EOM and BC, one can find out the independent degrees of freedoms in the multibrane setup. Note that it would be necessary to put all the permitted fluctuations into the metric.   
\begin{itemize}
\item[(1)] We should first examine the massless modes $\Yfund f(x) =0$  (without the GW field $\phi =0$) given by Eq.({\ref{eom}}) and Eq.(\ref{orth1}):
\beq
F'' - A' G' &=&   \frac{\kappa^2 }{3}
\sum_i \lambda_i  G \delta(y-y_i)  \\
F' - A' G &=& 0 \,\label{dec}
\eeq
Taking a second differentiation of Eq.({\ref{dec}}), one can see that  only the BC $\left[ A'   \right] |_i = \frac{\kappa^2}{3} \lambda_i$ 
is required to make the above two equations agree.  Most importantly,  one can immediately infer from Eq.(\ref{dec})  that $G(y)$ is continuous in the $y$-coordinate. 

Combining  the  two orthogonal conditions Eq.(\ref{orth1}) and Eq.(\ref{orth2}), one can derive the radion profile:. 
\begin{equation}  
F(y)=   
\begin{cases}    
c_1 \, e^{2A} +  k_1 \,  \epsilon'(y) \, e^{-2 A}  &  ,\, 0 < y < r \\[0.3cm]   
c_2 \, e^{2A} +  k_2 \,  \epsilon'(y) \, e^{-2 A}  &  , \, r < y < L   
\end{cases}     \, \label{FP}
\end{equation}
Note that the second term in the above equation is only half in the coefficient compared with~\cite{Kogan:2001qx}. We can calculate the related field using $G = F'/A'  $.
By imposing  the continuity conditions $F(r -\varepsilon) = F(r + \varepsilon)$ and $G(r -\varepsilon) = G(r + \varepsilon)$, one can determine  the junction condition: 
\beq
\left[ \epsilon''(y)  \right] \Large{|}_{y = r} &=& 4 (c_1- c_2) e^{ 4 A} 
= 4 \left( k_2 -k_1\right)\epsilon'  \,. \label{junction}
\eeq
as implicated by  the second orthogonal equation (\ref{orth2}) for $[A']\Large{|}_{y=r} = k_2 -k_1$. According to Eq.(\ref{junction}), one gets a class of solutions and it is the property of $[\epsilon''(y)]\Large{|}_{y=r}$ that determines whether $\epsilon'(r)$ is nonzero. 
However  since  $\epsilon'(y)$ can be arbitrary  away from the branes,  one can always tune $[\epsilon''(y)]\Large{|}_{y=r} =0$ to achieve  $\epsilon'(\pm r) = 0$ (equivalent to  $c_1 = c_2$),   same as $\epsilon'(0) = \epsilon'(L) = 0$ at the fixed points. For completeness, another EOM  can be  derived from $R_{55} = \kappa^2 \tilde{T}_{55}$. In the massless limit, the radion wave-function  obeys:
\beq
 && F'' f(x) - A' \left( G' + 2 F'\right) f(x) = \frac{\kappa^2}{3} G f(x) V(\phi_0)  \nonumber \\
 &&  + \frac{\kappa^2}{3} \sum_i  \lambda_i G f(x) \delta(y-y_i) 
\eeq
with $V(\phi_0) =- \frac{6}{\kappa^2} A'^2$. After simple algebra, the above equation is simplified to be $-2A' \left(F'- A' G \right)=0$, trivially satisfied due to Eq.(\ref{dec}).This confirms  there is no extra constraint for  $\epsilon'(y)$ in the bulk but its boundary values $\epsilon'(y) \Large{|}_{y =\{ 0, \pm r, L\}} $  are gauge invariant. Thus  we deduce  that $\epsilon'(y)$ is  a redundant  degree of freedom unless its bulk value is fully irrelevant and  a unique radion associated with  IR brane is the legitimate perturbation in the AdS$_5$ metric.  For $c_1= c_2$ and $\epsilon'(y) =0$, Eq.(\ref{FP}) reproduces the familiar radion solution  $G = 2 F  =  c \, e^{2 A}$ derived in the paper~\cite{Charmousis:1999rg}. 

\item[(2)] We will  look into the EOM (\ref{eom}) for a massive radion  by applying the GW stabilization.  From the  orthogonal equation~(\ref{orth2}),  the gauge fixing $G = 2F$ leads to  $\epsilon'(y) =0$ or $\epsilon'(y) \sim e^{4 A}$.  Supplemented with  the BC $\epsilon'(0)= \epsilon'(L)=\epsilon'(\pm r)  = 0$,   one can see that $\epsilon'(y)$ must be identically zero in  the two regions. Hence the EOM is simplified to be~\cite{Csaki:2000zn}:
\beq
e^{2A}F \Yfund f(x) + \left( F'' - 2 A' F' \right) f(x) =\frac{2}{3} \kappa^2 \phi_0' \varphi' \label{eom1} 
\eeq

\end{itemize}

A few properties related to the Lagrangian expansion in Eq.(\ref{Act}) are commented in order. After applying the EOM in the bulk, the tadpole term in the 4D effective Lagrangian with $\phi =0$ is calculated to be (see Appendix\ref{Appendix2}):
\beq
- \mathcal{L}_{tad} &= & \frac{4}{ \kappa^2} \int_{-L}^L d y e^{-4A}  \left( G -4 F  \right) A' (y)^2 f(x)    \nonumber \\
& +  &  \frac{4}{3}  \int_{-L}^{L} d y e^{-4 A} \sum_i  \lambda_i F f(x)   \delta(y-y_i)  \,. \label{tad}
\eeq
Using Eq.(\ref{dec}),  it is easy to verify that the tadpole term vanishes  with the brane potentials in Eq.(\ref{brane1}-\ref{brane2}). Note that the brane term at $y=-r$ due to the $Z_2$ orbifold symmetry is necessary for such cancellation.

The 4D Lagrangian at the leading order are the  kinetic terms. For the radion  one gets:~{\footnote{Due to  the conformally flat property of AdS$_{5}$, the first term of the radion kinetic term in Eq.(\ref{kin}) can be derived  from the Fierz-Pauli Lagrangian  in a straightforward manner, by replacing $h_{\mu \nu} \to - 2 F f(x) \eta_{\mu \nu}$ , $h_{55} \to 2 G \eta_{55}$ and $h \to 2 \left(G -4 F\right) f(x)$ in the Lagrangian of  ${\cal L}_{FP}= \frac{1}{2} \partial_\nu h_{\mu \alpha} \, \partial^\alpha    
h^{\mu \nu} -   \frac{1}{4} \partial_\mu h_{\alpha \beta} \, \partial^\mu h^{\alpha \beta}    
- \frac{1}{2} \partial_\alpha h \, \partial_\beta h^{\alpha \beta} +\frac{1}{4}    
\partial_\alpha h \, \partial^\alpha h $. The detail to derive the second term can be found in the paper \cite{Cai:2022geu}. }}  
\beq
\mathcal{L}_{kin} &=&- \frac{3}{\kappa^2} \partial^\mu f(x) \partial_\mu f(x)  \Big[  e^{-2 A}F\left( F -  G\right)  \nonumber \\ &-&  e^{-4 A}  \Big(F' -A' G -\frac{\kappa^2}{3} \phi'_0 \varphi \Big) \epsilon'  \Big ] \,, \label{kin}
\eeq
where  the second term in the bracket  drops out due to Eq.(\ref{orth1}). Before  stabilization ($\phi_0=0$), using $F' = A' G$, the coefficient of the kinetic term  can be rewritten as:
\beq
 \frac{3}{2 \kappa^2} \int_{-L}^{L} dy \frac{1}{A'} \frac{d \left(e^{-2 A} F^2 \right)}{dy} \,.
\eeq 
For $A' =$ {\it Constant},  this  only depends on the boundary values of $F(y)$. But after  stabilization, because of the fact  $F -  \frac{A'\epsilon'}{ e^{2A}} \not \propto e^{2 A} $ and $A' \neq$ {\it Constant} in each region, the normalization from Eq.(\ref{kin}) becomes  dependent on the bulk value of $\epsilon' (y)$. This indicates that $\epsilon' (y)$  must  be zero after the corresponding symmetry  is broken. It is not suitable to use  $\epsilon' (r)$ to create  another degree of freedom.

For clarity,  we remarks on the second solution in the paper~\cite{Kogan:2001qx},  obtained by  imposing the nonmixing condition  for the kinetic terms of 2 radions  in Eq.(\ref{kin}) (i.e. with  $F= F_1 + F_2 $ and $G = G_1+ G_2$):
 \beq
&&  \int_{-L}^L dy  \left(2 e^{-2A} \left[F_2 - \frac{A'\epsilon_2'}{ e^{2A}}\right] F_1   +  e^{-4A} F_1   \left[\epsilon_2'' - 2 A' \epsilon_2' \right] \right)  \nonumber \\ && \qquad  =  \, 0  \quad \quad  (\mbox{for } ~ \epsilon_1' =0 )  \,\label{nomix} 
 \eeq
For  $\phi_0 =0$ and $F_1 = c\, e^{2 A}$,  the second term equals $\int_{-L}^L dy \left[ e^{-4A} F_1 \epsilon_2' \right]'$  that precisely vanishes. Evaluated with  $F_2$ in the general expression of  Eq.(\ref{FP}),  this directly fixes the ratio:
  \beq
  \frac{c_2}{c_1} \simeq  -\frac{k_2}{k_1} e^{-2 k_2 (L-r)}  , ~\mbox{for}~ k_1 r \,, k_2 (L-r) \gg 1  \,   \label{ratio} 
  \eeq
indicating another zero mode  with  a spike at $y = r$  generated by  a nonzero $\epsilon_2'(r)$ (for $k_2 \to k_1$, $\epsilon'_2(r) \to \infty$). However after stabilization,  the second term in Eq.(\ref{nomix}) cannot be organized  into a  total differentiation anymore,  as a result  the  arbitrary bulk value of $\epsilon_2'(y)$  will enter.  This is also true  if $F_1$  holds a $\epsilon_1'$ part. Thus this unusual solution has to  be abandoned. 

Based on the analysis of degree of freedom,  it is valid to take the intermediate brane  as  {\it rigid} in a static solution,  with its location  fixed by the boundary of unequal cosmology constants.  In  this setup,  the  $r$ parameter is traded with the scalar VEV  at the $y =  r$ brane.  We now pursue  to stabilize the radion  associated with the  separation between the UV and IR branes.  The effective potential of radion is given by integrating over the fifth dimension:
 \beq 
& & V_{eff} (y_L) \simeq 2 \int_0^{y_L} dy ~ e^{- 4 A } \left[\frac{1}{2}   
(\partial_5 \phi_0)^2 + V(\phi_0) \right] \nonumber \\
& & +  e^{- 4 A(y_L)}  \lambda_{L} (\phi_0(y_L)) \, \label{pot} 
\eeq 
where  the VEV of the GW bulk scalar is  $\phi_0 = \phi_P e^{-u y}$ with $\phi_P$ denoting the UV brane value. Then the first derivative of the potential is:
 \beq
\frac{\partial V_{eff}}{\partial y_L} {\Big |}_{y_L= L} =  \frac{\kappa^2}{3} e^{- 4 A (L)} W (\phi_0(L))^2 \, = \, 0 .
\eeq
and the extrema gives  $L = \frac{1}{u} \log \frac{\kappa \phi_P}{\sqrt{6 k_2/u}}$.

The mass of radion in RS1 was calculated in~\cite{Csaki:2000zn, Tanaka:2000er}. Following the procedure,
we expand the background metric $A(y)$ and the  radion wave function $Q(y)$  in a measure of the backreaction:
\beq
Q &=& \begin{cases}  e^{2k_1 |y| }\left[1 + l^2 f_1(y) \right]&  ,0 < y < r  \\[0.3cm]  
e^{2k_2 |y| + 2 r(k_1 - k_2)} \left[1 + l^2 f_2(y) \right] & ,r < y < L   
\end{cases}   
\nonumber \\
\mbox{and}  \nonumber 
\\
A & =& \begin{cases} k_1 |y| + \frac{l^2}{6} e^{-2u |y|}  &  ,0 < y < r  \\[0.3cm]  
 k_2  |y| + (k_1 -k_2 ) r + \frac{l^2}{6} e^{-2u |y|} & ,r < y < L   
\end{cases}   
\eeq
with $l = \kappa \phi_P/\sqrt{2}$  and  the mass  parametrized as $m^2 = \tilde{m}^2 l^2 $. The $\varphi'$ in the  EOM  (\ref{eom1}) can be eliminated using Eq.(\ref{orth1}). At the zeroth order the EOM yields the  massless case without back-reaction. Expanding to the  $l^2$ order, one derives:
\beq
f''_1 + 2(k_1 +u) f'_1 &=& -\tilde{m}^2 e^{2k_1 y} -\frac{4 (k_1 -u )u }{3} e^{-2 u y} \\
f''_2 + 2(k_2 +u) f'_2 &=& -\tilde{m}^2 e^{2k_2 y + 2(k_1-k_2)r}  -\frac{4  (k_2 -u ) u}{3} e^{-2 u y} \nonumber
\eeq
In the limit of a stiff brane potential, namely $\gamma_{\pm} \to \infty$, the BC reduces to be $(Q' - 2 A' Q)\Large{|}_{y = \{0, L\}} =0 $, hence gives $(f_{1,2}' +\frac{2}{3} u e^{-2 u y })\Large{|}_{y = \{0, L\}} =0$. At the $y= r$ brane we impose the continuous BC from the {\it jump} matching~\footnote{In general, the mass parameter in the superpotential  may  be set to be unequal ($u_1\neq u_2$) in two spatial regions. Firstly, the brane tension gets a shift $ \frac{1}{2} (u_1 - u_2 ) \phi_0^2 $ and the BC of  radion profile  at $y=r $ alters to be: $ u_2 f_1'(r- \varepsilon) - u_1 f_2' ( r+ \varepsilon) = 0$ for such case. In addition, to be compatible with the  scalar $\phi$ BC,  i.e.  $[\varphi'  - 2 F \phi'_0]|_{y=r} = 0$, one needs to impose the constraint $(u_2 - u_1) \left( \phi_0 \varphi - \frac{3}{\kappa^2} \frac{ e^{2A} }{u_1 u_2} \Box F \right) \big{|}_{y =r} =0$, that is trivially satisfied if $u_1 = u_2$.}:
\beq
f_1'(r- \varepsilon) = f_2' ( r+ \varepsilon) 
\eeq
After a lengthy calculation,  the mass of radion is determined by BC at the $y= L$ brane: 
\beq
 m^2  &=&   \frac{ 4 u^2  (2 k_2 + u)l^2}{3 k_2} e^{-2 \left[(k_2+ u) L + ( k_1 - k_2) r \right] }  \nonumber \\ & -& C \, l^2 \, e^{- 2 \left[(2 k_2  + u ) L + 2 ( k_1 - k_2) r \right]}  \,, \label{mass}
\eeq
with the constant  fixed by the other two BCs, 
\beq
C &\simeq & \frac{4u^2 \left( 2 k_2 + u \right)}{3 k_1 k_2} \Big[\left(k_2 - k_1 \right) e^{2 k_1 r} - k_2 \Big]\,, \nonumber \\ && (\mbox{for}~~ 0 \ll r \ll L) \,.
\eeq
Thus the last term in the radion mass (\ref{mass}) is  negligible due to the  large suppression from a  warped factor. 

Finally we briefly discuss the  cosmological expansion rate by taking the  metric  to  be of a time evolution form:
\beq
&& ds^2 = n(t, y)^2 dt^2 - a(t, y)^2  dx^2 - b(t, y)^2 dy^2  
\\ \nonumber \\
&& a(t, y)  = a_0(t) e^{-A} (1+ \delta a) \,, \, n(t, y) = e^{-A} (1+ \delta n) \nonumber \\
&&  b(t, y) = 1+ \delta b \nonumber
\eeq
The perturbations $(\delta a, \delta n , \delta b)$ are caused by adding the matter densities. We will inspect the $G_{55}$ equation using the following ansatz:
\beq
G_{55} &=&  3\left (  \frac{a'}{a} \left( \frac{a'}{a} + \frac{n'}{n} \right) -\frac{b^2}{n^2} \left(\frac{\dot a^2}{a^2} -  \frac{\dot a \, \dot n }{a \, n}  + \frac{\ddot  a}{a} \right) \right) \nonumber  \\
T_{55} &=& \frac{1}{2} \phi'^2_0 - b^2 \, V(\phi_0)
\eeq
Taking the jump of $G_{55} = \kappa^2 T_{55} $  at the  $y =r $ brane where the reflection symmetry  is not operative, 
we obtain at the leading order:
\beq
&&  3 \Big( \left[A' \right] \big{|}_{ r}  \langle 3  \delta a'  + \delta n' \rangle  + \langle A' \rangle   \left[ 3 \delta a' + \delta n' \right] \big{|}_{r } \Big) \nonumber\\ && =  \Big[6  A'^2  - \frac{\kappa^2}{2} \phi'^2_0 + \kappa^2 V(\phi_0) \Big] \Big{|}_{ r } - 12 \left[ A'^2 \right] \big{|}_{r} \delta b \label{G55a}
\eeq
where  the  first quantity on the right side  vanishes due to Eq.(\ref{BG2}) and  $\langle \delta a'(r) \rangle$ represents the average of  $\frac{1}{2}\left( \delta a'(r + \varepsilon ) +  \delta a'(r - \varepsilon ) \right)|_{\varepsilon \to 0}$. The jump equations for $\delta a'$ and $\delta n'$ are given in \cite{Binetruy:1999ut, Csaki:1999mp}:
\beq
[\delta a' ]|_{r} = -\frac{\kappa^2}{3} \left( \rho + \lambda_r \delta b \right) , 
[\delta n' ]|_{r} = \frac{\kappa^2}{3} \left( 3p + 2\rho - \lambda_r \delta b \right)  \nonumber \\ \label{delta1}
\eeq
with $\rho$ and $p$ being  the matter density and pressure at the $y=r$ brane. By replacing Eq.(\ref{delta1}) in Eq.(\ref{G55a}), one arrives that,  
\beq
\langle 3  \delta a'  + \delta n' \rangle  = \frac{\kappa^2}{6} \frac{k_2+k_1}{ k_2 - k_1 } \left(   \rho - 3 p \right)  - 4 \langle A'\rangle \delta b \label{delta2} 
\eeq
Note that there is no singularity at $k_1 = k_2$ in the above equation since the density $\rho$ along with the intermediate brane will disappear in that limit.  Then we will average the $G_{55} = \kappa^2 T_{55} $ with respect to the $y=r$ brane by keeping only the linear term of $\rho$:
\beq
  \left( \frac{\dot a_0}{a_0} \right)^2  + \frac{\ddot a_0}{a_0}    &=& e^{-2A} \Big ( \frac{\kappa^2 \left[  A'\right]|_{r} }{12} ( \rho - 3p) \nonumber \\
&-& \langle A' \rangle   \langle 3  \delta a'  + \delta n' \rangle    \nonumber \\ &-& \Big(4 \langle A' \rangle^2 - \frac{\kappa^2}{3} \phi'^2_0 \Big) \delta b  \Big) \, \label{FRW}
\eeq
Substituting  the $\langle 3  \delta a'  + \delta n' \rangle $ term  in  Eq(\ref{FRW}) with Eq.(\ref{delta2}), we derive  the FRW equation near the intermediate brane to be:
\beq
 \left( \frac{\dot a_0}{a_0} \right)^2  + \frac{\ddot a_0}{a_0}  &=& \frac{  e^{-2 A}  }{3 }  \frac{ \kappa^2 k_1 k_2   }{k_1 - k_2}  \left( \rho - 3 p \right)  \nonumber \\ &+&  \frac{e^{-2 A}}{3}   \kappa^2\phi'^2_0   \delta b   \,. \label{FRW2}
\eeq
For  consistency with  late time cosmology,  one expects $\left( \frac{\dot a_0}{a_0} \right)^2  + \frac{\ddot a_0}{a_0} \sim \frac{1}{6 M_{Pl}^2} \sum_i (\rho_i - 3 p_i)$ where $(\rho_i, p_i)$ are the physically measured quantities on each brane.   Although, without knowing $\delta b$, it is not possible to make a robust prediction,  nonetheless,  Eq.(\ref{FRW2}) is a constraint that the multibrane system needs to satisfy.    

In summary,  in this letter we derived  two orthogonal conditions to decouple the transverse  graviton  from the modulus field in a multibrane RS model.   By solving the  linearized Einstein's equation we find out that  the  perturbation $ \epsilon(y) \partial_\mu \partial_\nu f(x) $ merely plays the role of gauge fixing for the radion field and can not be used to create a new excitation  in the presence of radius  stabilization. The intermediate brane originates from the necessity to match the jump condition of  the background metric. To get a GW-like minima from the effective potential,  it is reasonable to  assume that the  nonfixed point brane is (quasi) {\it{rigid}},  with a time evolution due to matter densities.   Under such rigid assumption,  the  radius of IR brane can actually  be promoted to be a dynamic field. Instead the  location of intermediate brane  is purely  a parameter that  signals the discontinuity of bulk cosmology constants.  Hence one can anticipate that the stabilization of the IR radius in the multibrane model  is similar to the RS1,   with the minima affected by the unequal curvatures. After applying the GW mechanism, we show that  with a small back-reaction, the radion mass  is well below  the cut off  scale  of IR brane, that is consistent with the NDA argument from  the AdS/CFT correspondence~\cite{Maldacena:1997re, Witten:1998qj, Gubser:1998bc}.  A favorable property is  that the radion and its Kaluza-Klein towers after the proposed stabilization are orthogonal in the limit of  stiff brane potentials.

\section*{Acknowledgments}
H.C. \ is supported by the National Research Foundation of Korea (NRF) grant funded by the Korea government (MEST) (No. NRF-2021R1A2C1005615).

\raggedright
\bibliography{radion}
\onecolumngrid
\input{appendix}

\end{document}

%% file: appendix.tex
\appendix
\section{Einstein's equation in the multibrane model}\label{Appendix}

The Einstein's equation $R_{MN} = \kappa^2 \tilde{T}_{MN} \equiv \kappa^2 \left(T_{MN} -\frac{1}{3} g_{MN} T^a_a\right)$  determines the dynamics of  metric fields.  We derived the components for the Ricci tensor and the energy-momentum tensor in a  RS-like model. 

\beq
R_{\mu \nu} &=& R_{\mu \nu}^{(h)} + R_{\mu \nu}^{(f)} \, \label{Ruv} 
 \eeq
 
 \beq
R_{\mu \nu}^{(h)} &=& \frac{1}{2} \left( \partial_\mu \partial_\lambda   
h^{\lambda}_{\nu}  +  \partial_\nu  \partial_\lambda h^{\lambda}_{\mu}    
-  \Yfund h_{\mu \nu} -  \partial_\mu  \partial_\nu h  \right) \nonumber \\
&+& \frac{1}{2} e^{-2 A} \left( \partial_5^2 h _{\mu \nu} - 4 A'  \partial_5 h_{\mu \nu}  \right) \nonumber \\
&+& \left[ 4 A'^2 - A'' \right] e^{-2 A} h_{\mu \nu} - \frac{1}{2} e^{-2 A} \eta_{\mu \nu}A' \partial_5 h \, \label{Ruv1}
\eeq

\beq
R_{\mu \nu}^{(f)} &=& e^{-2 A} \eta_{\mu \nu} \left[  \left[ 4 A'^2 - A''  \right] \left( 1- 2(G+F) \right) + A'  \left( 8  F' +  G' \right)  -   F''  \right] f(x)  \nonumber \\
&+& e^{-2 A} \left[ 2 \left[  4 A'^2 - A'' \right]   \epsilon (y)   +  \epsilon''(y) - 4 A' \epsilon'(y) \right] \partial_\mu \partial_\nu f (x)     \nonumber \\
&+&  (2 F -G) \partial_\mu \partial_\nu  f(x) + \eta_{\mu \nu} F \Yfund  f(x) - e^{-2 A} A' \epsilon'(y) \eta_{\mu \nu} \Yfund f(x)  \, \label{Ruv2}
\eeq

\beq
R_{\mu 5} &=& -\frac{1}{2} \left( \partial_{\mu} \partial_5 h - \partial_\alpha \partial_5 h^{\alpha}_{\mu} \right) + 3 \left(  F'   - A'   G \right)  \partial_\mu f (x)  \, \label{Ru5}
\\ \nonumber \\ 
R_{55} &=&4 \left(A''- A'^2\right)- \frac{1}{2} \left( \partial_5^2 h  - 2 A' \partial_5 h  \right) - \left[ \epsilon'' (y)  - 2 A'  \epsilon' (y)  \right] \Yfund f(x) \nonumber \\
&+&  e^{2 A}  G \Yfund f (x) + 4  F'' f (x)  -4 A' \left[ G' + 2 F' \right] f (x)  \,\label{R55}
\eeq

 \beq
 \tilde{T}_{\mu \nu} &=& -\frac{2}{3} e^{-2A} \left[ \eta_{\mu \nu} 
\left( V(\phi_0) + V'(\phi_0) \varphi -2  V(\phi_0) F f(x) \right) + V(\phi_0) h_{\mu \nu} \right]  \nonumber \\ 
& -& \frac{1}{3} e^{-2A}
\sum_i \left[  \eta_{\mu \nu} \left(\frac{\partial \lambda_i(\phi_0)}{\partial \phi} \varphi
+ \lambda_i(\phi_0) \left[1-(2 F+ G) f (x) \right] \right) + h_{\mu \nu}   \lambda_i (\phi_0) \right] \delta(y-y_i) \nonumber \\
&-&  \frac{4}{3}  e^{-2A} \epsilon(y) \partial_\mu \partial_\nu f (x) V(\phi_0)   - \frac{2}{3} e^{-2A}   \partial_\mu \partial_\nu f (x)  \sum_i \lambda_i (\phi_0) \epsilon( y) \delta (y -y_i)   \label{Tuv}
 \eeq
 \beq
  \tilde{T}_{\mu 5} &=& \phi_0' \partial_\mu \varphi
 \\  \nonumber \\
 \tilde{T}_{55}^{(h)} &=&   \phi'_0 \left(\phi_0'+ 2 \varphi' \right)+  \frac{2}{3}  V(\phi_0) \left[1+ 2 G f(x) \right] + \frac{2}{3}V'(\phi_0) \varphi \nonumber \\ &+& \frac{4}{3}  \sum_i \left( \lambda_i \left[ 1 + G f(x) \right] + \frac{\partial \lambda_i(\phi_0)}{\partial \phi} \varphi \right)\delta (y -y_i)    \label{T55}
 \eeq
 
From the derivation,  we  can just extract out  the graviton terms. In this way, the Einstein's equation gives the equation of motion (EOM)  for 5d graviton  subject to the  transverse and traceless gauge fixing:  
 \beq
  e^{2 A} \partial_5 \left( e^{-4 A} \partial_5 h_{\mu \nu} \right) &=& \Yfund h_{\mu \nu} \label{GE}   \\
  2 R_{\mu 5} \supset - \partial_5 \left( \partial_\mu h -\partial_\nu  h^{\nu}_{\mu}\right) &= & 0 \nonumber \\
   2  R_{5 5} \supset - \left( \partial_5^2 h - 2 A'  \partial_5 h \right) &=& 0 \nonumber
 \eeq
 where  the  background equation~(\ref{BG1}) is applied for simplification. Note that the boundary conditions for Eq.(\ref{GE}) are obtained by matching the singular terms, i.e. $\partial_5 h_{\mu \nu}\Large{|}_{y=\{ 0, L\}} =0$ and $[\partial_5 h_{\mu \nu}]\Large{|}_{y=r} = 0$
 
 \section{The absence of tadpole terms}\label{Appendix2}
Without the GW bulk scalar ($\phi=0$), expanding the 5d action Eq.(\ref{Act}) to the linear order of metric perturbations, we can get the tadpole terms:
\beq
\mathcal{L}_{tad} &= & \frac{1}{2 \kappa^2} \int d y \, 8\,  e^{-4A}  \left(  \left[F''-A' G' \right] - 2 A'' G   - 5 A' \left[ F'- A' G\right] \right) f(x) \nonumber \\
& - &  \frac{1}{2 \kappa^2} \int d y  e^{-4A} \left[ G -4 F  \right] f(x) \left( 20 A'^2 - 8 A''\right) \nonumber \\
& - &   \int dy e^{-4A}  \left(\left[G -4 F \right] f(x) V -4 F f(x) \sum_i  \lambda_i \delta(y-y_i) \right) \,. \label{tad0}
\eeq
From  the BG equations~(\ref{BG1}) and (\ref{BG2}),  we can identify:
\beq
V  &=& - \frac{6}{\kappa^2} A'^2     \nonumber \\ 
A''  &=&  \frac{\kappa^2}{3} \sum_i \lambda_i \delta(y-y_i) \label{BG3}
\eeq
For Eq.(\ref{tad0}), we can first apply the EOM (\ref{dec}) of the massless mode, then substitute Eq.(\ref{BG3}) into Eq.(\ref{tad0}).  The tadpole terms are simplified to be:
\beq
- \mathcal{L}_{tad} &= & \frac{4}{ \kappa^2} \int d y  e^{-4A}   \left( G -4 F  \right) f(x)  A'^2   \nonumber \\ 
& +  &  \frac{4}{3}  \int d y e^{-4 A} \sum_i  \lambda_i F f(x)   \delta(y-y_i) 
\eeq
The above ansatz is exactly Eq.(\ref{tad}) in the main text. We can apply for a further transformation using $G = F'/A'$, and  this gives:
\beq
- \mathcal{L}_{tad} &= & \frac{4}{ \kappa^2} \int_{-L}^L d y  \frac{d (e^{-4A} F) }{dy}  A'  f(x)    \nonumber \\
& +  &  \frac{4}{3}  \int_{-L}^{L} d y e^{-4 A} \sum_i  \lambda_i F f(x)   \delta(y-y_i) 
\eeq
Therefore  with $\lambda_+ = \frac{6 k_1}{\kappa^2}$, $\lambda_- = - \frac{6 k_2}{\kappa^2}$ and $ \lambda_{\pm r}  = \frac{3 (k_2- k_1)}{ \kappa^2}$ in the massless limit, the tadpole terms  vanish  in the 4D effective Lagrangian.